# Momentum Conserved Ultrafast Charge Transfer Dynamics of Interlayer Excitons in vdW Heterostructures

*Pranjal Kumar Gogoi,\*,† Yung-Chang Lin,† Ryosuke Senga,† Hannu-Pekka Komsa,‡ Swee Liang Wong,¶ Dongzhi Chi,¶ Arkady V. Krasheninnikov,‡,§ Lain-Jong Li,∥ Mark B. H. Breese, ⊥,# Steven J. Pennycook,@ Andrew T. S. Wee,⊥ and Kazu Suenaga,\*,†*

†National Institute of Advanced Industrial Science and Technology (AIST), AIST Central 5, Tsukuba 305 - 8565, Japan

‡Department of Applied Physics, Aalto University, P.O. Box 11100, 00076 Aalto, Finland

¶Institute of Materials Research and Engineering, Agency for Science, Technology and Research, 2 Fusionopolis Way, #08 - 03 Innovis, Singapore 138634

§Helmholtz-Zentrum Dresden-Rossendorf, Institute of Ion Beam Physics and Materials Research, 01328 Dresden, Germany

∥Physical Science and Engineering Division, King Abdullah University of Science and Technology (KAUST), Thuwal, Saudi Arabia, 23955

⊥Department of Physics, Faculty of Science, National University of Singapore, Singapore 117542

#Singapore Synchrotron Light Source, National University of Singapore, 5 Research Link; Singapore 117603

@Department of Materials Science & Engineering, National University of Singapore, 9 Engineering Drive 1, Singapore 117575

CORRESPONDING AUTHORS

*pranjalkumar-gogoi@aist.go.jp

†suenaga-kazu@aist.go.jp

ABBREVIATIONS

vdW, TMDC, 2D, hMW, EEL, STEM

ABSTRACT

Heterostructures comprising van der Waals (vdW) stacked transition metal dichalcogenide (TMDC) monolayers are a fascinating class of two-dimensional (2D) materials with unique properties. The presence of interlayer excitons, where the electron and the hole remain spatially separated in the two layers due to ultrafast charge transfer, is an intriguing feature of these heterostructures. Inevitably, the efficiency of 2D heterostructure devices is critically dependent on the charge transfer dynamics. However, the role of the relative rotation angle of the constituent layers on this charge transfer dynamics is hitherto unknown. Here, we investigate $MoS_2$/$WSe_2$ vdW heterostructures (hMWs) using monochromated low-loss electron energy loss (EEL) spectroscopy combined with aberration-corrected scanning transmission electron microscopy (STEM), and report that momentum conservation is a critical factor in the charge transfer dynamics of TMDC vdW heterostructures. The low-loss EEL spectra of the heterostructures with various rotation angles reveal that the charge transfer rate can be about one order-of-magnitude faster in the aligned (or anti-aligned) case than the misaligned cases. These results provide a deeper insight into the role of the fundamental principle of momentum conservation in 2D vdW heterostructure charge transfer dynamics.

The advent of 2D materials has been an inflection point both for recent condensed matter physics as well as material science research.[1] Starting with the successful isolation of graphene and subsequent investigations on single and few layer TMDCs, the discovery of a plethora of fundamental phenomena as well as observation of novel properties suitable for diverse potential technological applications, particularly in fields like ultrafast electronics, optoelectronics, have established 2D materials as one of the hottest current research fields.[2-4] One recent topic of focus in this field has been vdW heterostructures made of two or more single atomic layers, which show compelling new properties beyond those of the constituent layers.[5-16] While the fabrication of heterostructures of high mobility quantum superlattice systems require sophisticated techniques such as molecular beam epitaxy, 2D vdW heterostructures could be made using exfoliation or chemical vapor deposition (CVD) growth of individual layers and then stacking them manually. Direct growth of vertical and lateral heterostructures by CVD with atomically abrupt boundaries have been also demonstrated.[13, 17] vdW heterostructures comprising constituents such as two layers of graphene,[18, 19] graphene and boron nitride,[20] and various combinations of 2D TMDC materials[6, 21-24] have been reported. These heterostructures show novel properties such as Hofstadter butterfly states, photo-induced doping, and interlayer excitons with ultrafast charge transfer. These heterostructures possess a new degree-of-freedom, which is expected to critically determine their properties and functionality, i.e., the relative orientation of the individual layers. The fundamental role of the relative orientation of the graphene layers in bilayer graphene Moiré superlattices has been successfully demonstrated recently in the seminal result of observation of superconductivity and correlated insulator behavior at the so-called magic angles.[25, 26]

In 2D TMDCs, the individual single layers exhibit extremely strong many-body effects in the form of different exciton species with binding energies in the range of ~0.4 to 1 eV, due to reduced screening.[3, 27-29] Heterostructures made of $MoS_2$ and $WSe_2$ are known to form a type-II staggered bandstructure.[11, 23] It has been demonstrated that these kind of heterostructures (e.g., $MoS_2/WSe_2$ and $MoS_2/WS_2$) reveal the presence of interlayer excitons which form after ultrafast charge transfer.[6, 12, 30, 31] The absorption of light occurs in the individual layers, and the holes and electrons move into the adjacent layer forming interlayer excitons, where the holes and electrons are spatially separated. These excitons thereafter emit radiatively as observed in photoluminescence experiments.[6, 7, 24]

The formation of interlayer excitons are due to ultrafast charge transfer at the time scale of few tens of femtoseconds, and this process competes with other relaxation channels and radiative recombination within the same layer.[12, 30, 32, 33] The ultrafast charge transfer process is critical for the manifestation of interlayer excitons, which has potential applications in ultrafast photodetectors, photovoltaics, and photocatalysis, and other light detecting and harvesting technologies. Energy and momentum conservation are expected to play a decisive role in the ultrafast charge transfer process and its efficiency. Studies on the fundamental conservation principles of charge transfer dynamics are still lacking, mainly due to the challenge of simultaneously determining the relative orientation of the layers and acquiring (optical excitation) spectra at high spatial resolution, eliminating the possibility of spurious results due to local inhomogeneities.[34, 35] The relative orientation of the layers incorporates the momentum match or mismatch of the electrons and holes pertaining to the optical excitation of these systems. We exploit this property to investigate how momentum conservation determines or influences the charge transfer rate in a $MoS_2/WSe_2$ vdW heterostructure.

We employ highly monochromated EEL spectroscopy in the low-loss energy range in an aberration-corrected STEM to determine the optical excitation response of hMWs as a function of the rotation angle between the layers. The low-loss EEL spectra (referred to as EEL spectra hereafter) depend on the complex dielectric function in general, and reflects important features such as the optical band-gap, exciton and interband transitions. The high energy resolution of the monochromated electron beam combined with the high spatial resolution of the aberration-corrected STEM provide unprecedented capability to determine both the optical response and relative orientation (from STEM high angle annular dark field (HAADF) images) of the layers with nanometer resolution. Our main observation is that for the aligned (0°) and anti-aligned (60°) cases, the interlayer scattering is approximately one order-of-magnitude faster than in the misaligned cases (e.g., 30°). The scattering rate is another manifestation of the interlayer charge transfer rate when the electrons and holes move to the adjacent layer. We also observe a small redshift of the absorption peaks in the heterostructure with respect to the peak positions of the individual layers, which we interpret as due to increased screening by the adjacent layer. This is dependent on the interlayer distance, which has local modulations from steric effects.

**Experimental procedure**

In this work, an aberration-corrected electron microscope operating in the STEM mode at 60 kV is used, as shown schematically in Figure 1(a).[36-38] Details of the experimental set-up can be found in the Methods section. The hMW samples are fabricated from individual single layers of $MoS_2$ and $WSe_2$ grown by CVD (see Methods). The $MoS_2$ layer is transferred onto the $WSe_2$ sample (Figure 1(b)). The $MoS_2/WSe_2$ bilayer is then transferred onto copper grids. One important aspect of these hMW samples is that the randomly-orientated individual single crystal triangles of $MoS_2$ are much smaller in size (~1-3 micron sides) than the triangular single crystals of $WSe_2$ (~10 micron sides). Because of this deliberate choice of sample sizes, there are many $MoS_2$ triangles with random orientations on a larger single crystalline $WSe_2$ triangle. This enables us to acquire spectra from a large number of areas with different relative orientations. In this work, we present EEL spectra for hMWs with 29 different orientation angles (see Supplementary Figure S3). Figure 1(c) and (d) show the STEM HAADF images from hMWs with 29° and 50° misorientations, respectively, as representative cases. The fast Fourier transform (FFT) patterns from these images are used to determine the rotation angles as illustrated in the schematics of Figure 1(e) and (f). The 0° (aligned) case represents the R stacking while 60° (anti-aligned) case represents the H stacking, which can be discriminated from the Moiré patterns in the STEM HAADF images (see Supplementary Figure S6).[39]

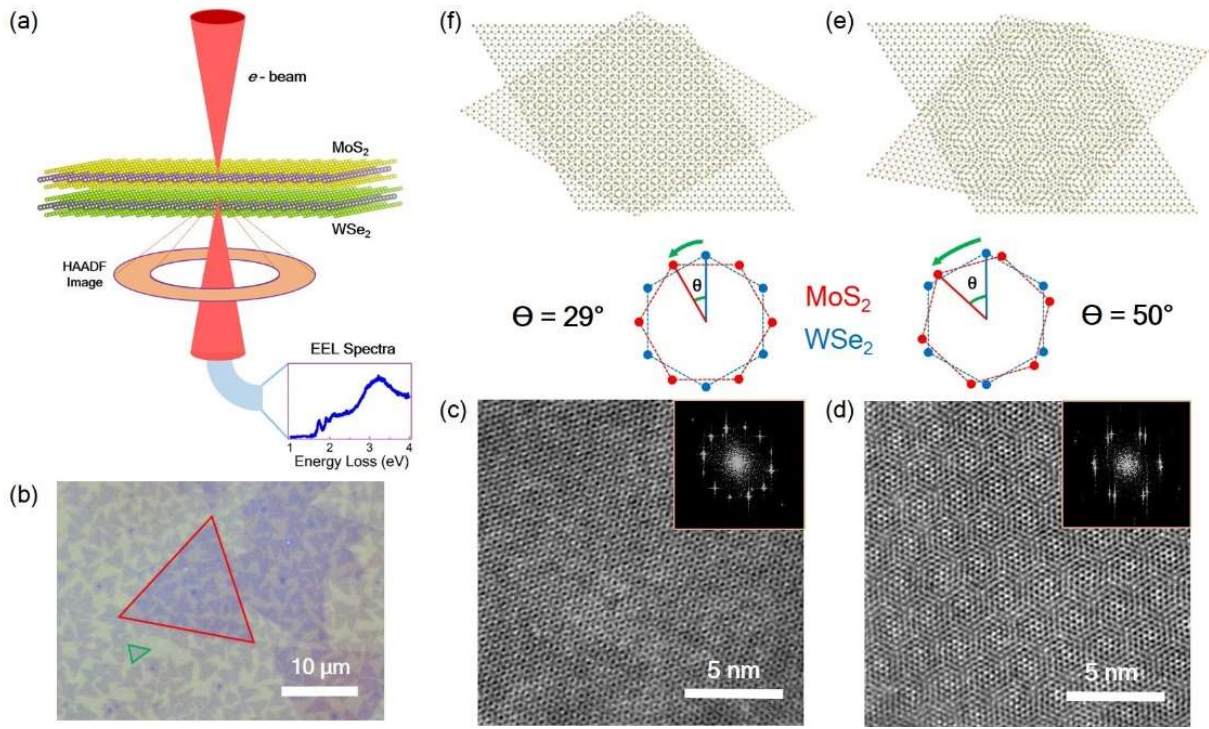

**Figure 1.** Schematic of the STEM-EEL spectroscopy experimental set-up and hMW sample images. (Anticlockwise from top left) (a) Electron beam converges on the sample plane and scans on the sample giving both high spatial resolution images and high energy resolution EEL spectra. (b) Optical image of a hMW sample. Bigger red triangle indicates one of the monolayer $WSe_2$ single crystals, while the smaller green triangle indicates one of the many monolayer $MoS_2$ single crystals. This optical image is taken after the transfer of the $MoS_2$ to the $WSe_2$ sample on substrate. (c) STEM HAADF image of the Moiré superlattice of hMW with 29°. The schematic of the two layers in this rotational alignment is shown in (f) above. (d) and (e) represent the case for hMW with 50°. The respective fast Fourier transform (FFT) patterns from the HAADF images are shown in the insets of (c) and (d).

## Results and discussion

### Layer rotation angle-dependent EEL spectra

In Figure 2(a), the measured EEL spectra is plotted for monolayer $MoS_2$, $WSe_2$, and hMWs with different relative orientations as representative cases after zero-loss subtraction using a standard power-law fit (see Supplementary Figure S1 & S2).[40] The loss-function is given by the imaginary part of $[-1/\epsilon(q,E)]$, where $\epsilon(q,E)$ is the dielectric response function, $q$ is momentum, and $E$ is energy. In the case of optical probes, $q$ is negligible, and hence we observe the signatures of dipole-allowed transitions in the optical dielectric function spectra, which can be mimicked in an EEL experiment performed in the TEM mode. In our case, the experiment is performed in the STEM mode which incorporates nonzero momentum changes to the charge carriers during excitations, due to the finite convergence and collection angles used. However, as can be seen in the monolayer EEL spectra of $MoS_2$ and $WSe_2$, the overall shape and peak positions are remarkably close to the optical case (see Supplementary Figure S8). In particular, the so called A, B, C peaks of $MoS_2$, and A, B, C, D peaks for $WSe_2$ observed in optical experiments and discussed in the literature are also found to be the most prominent features of the EEL spectra as can be seen in Figure 2(a).[12, 41, 42] There are a couple of small peaks after the B exciton peak in $MoS_2$ EEL spectra, which could be due to transitions with finite momentum change. Overall, the close resemblance of the EEL spectra acquired in the current STEM-EEL spectroscopy setting with that of optical absorption features (specifically, with $\epsilon_2$) emphasize the fact that these spectra could be used directly for understanding processes and extracting physically meaningful parameters as a first approximation, without invoking the complex and computationally challenging finite momentum change in the analysis. We take a similar approach and base our analysis and interpretation in this work on the EEL spectra and unravel some important results.

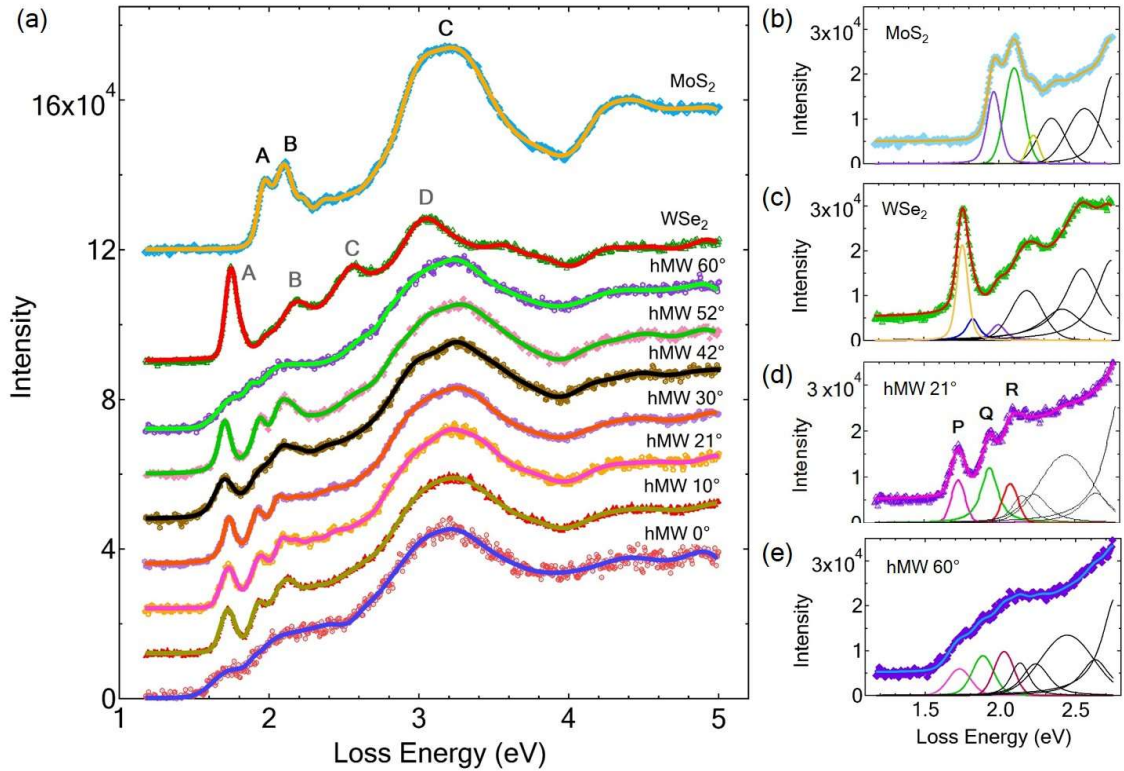

**Figure 2.** Relative orientation-dependent EEL spectra of hMW. (a) Comparison of the EEL spectra of monolayer $MoS_2$ and $WSe_2$ with that of aligned (ant-aligned) and misaligned hMW. (b) EEL spectrum of monolayer $MoS_2$ upto 2.75 eV, showing the Gaussian-Lorentzian oscillators used to fit the spectrum. (c), (d), and (f) represent the similar cases (as for (b)) for $WSe_2$ monolayer, hMW with 21˚, and hMW with 60˚, respectively, as representative cases.

For the hMWs, the EEL spectra is different from both the monolayer $MoS_2$ and $WSe_2$. However, as can be seen from Figure 2(a), particularly for the misaligned cases (~10˚-52˚), the spectra can be interpreted as originating from the effective dielectric response of the two layers together with weighted average contributions (see Supplementary Figure S7).[12] Three relatively well separated peaks (P, Q, and R as labelled in Figure 2(d)) are observed for the misaligned cases until ~2.4 eV in energy. More importantly, these peaks are strongly broadened for the aligned (0°) and anti-aligned (60°) cases. This observation is the key evidence of the enhancement of the charge transfer rate, as will be discussed later. For determining the peak shifts and broadening accurately for quantitative analysis, we perform fitting of the EEL spectra using multiple physically meaningful Gaussian-Lorentzian line-shapes corresponding to each structure (or peak) in the spectra. Since our focus is on the first three peaks and their lineshapes, we perform data fitting up to 2.75 eV for all the hMWs with 29 different orientation angles. The results from these fittings are used for subsequent analysis and discussion in this work. The representative cases of the fit results and the constituent line-shapes are shown in Figure 2(b)-(e) for $MoS_2$, $WSe_2$, and hMWs with 21° and 60° relative orientations, respectively.

**Redshift and broadening of heterostructure peaks**

In Figure 3(a), lineshapes of the A peak of $WSe_2$ and the P peak of hMW at 21° and 60° as representative cases are compared. Two distinct features are observed - there is a small redshift of the P peak with respect to the A peak of $WSe_2$, and the P peak is drastically broadened for the 60° case. The peak shift ($\Delta_1 = A - P$) is plotted as a function of the rotation angle in Figure 3(b). We note that variations of the A peak position is observed for even single layers of $WSe_2$ in different holes in the grid (see Supplementary Figure S5). This can be attributed to unintentional local doping variations and strain effects.[43, 44] To eliminate systematic errors and to accurately estimate the shift relevant to a particular sample location, $WSe_2$ and hMW EEL spectra are collected from every location. $\Delta_1$ is calculated from these spectra by subtracting the peak positions of $WSe_2$ and hMW from the same location. We observe that $\Delta_1$ is almost constant with no perceivable difference with rotation angle of the layers. For aligned (and ant-aligned) cases and close to these, the error bars are rather large due to the large broadening and hence no definitive conclusions can be drawn for those cases. However, for all other misaligned cases, the error bars are reasonable, and we estimate $\Delta_1$ as $22 \pm 10$ meV. We interpret this shift as due to increased screening by the adjacent layer as reported previously.[6, 12, 45] Importantly, the broadening of the P peak of hMW shows distinct rotation angle dependence (Figure 3(c)), where the broadening becomes approximately more than double for the aligned and anti-aligned cases in comparison to the misaligned cases. In Figure 3(d) the broadening of the A peak of WSe$_2$ and the P peak of hMW are shown for the aligned, anti-aligned, and misaligned cases.

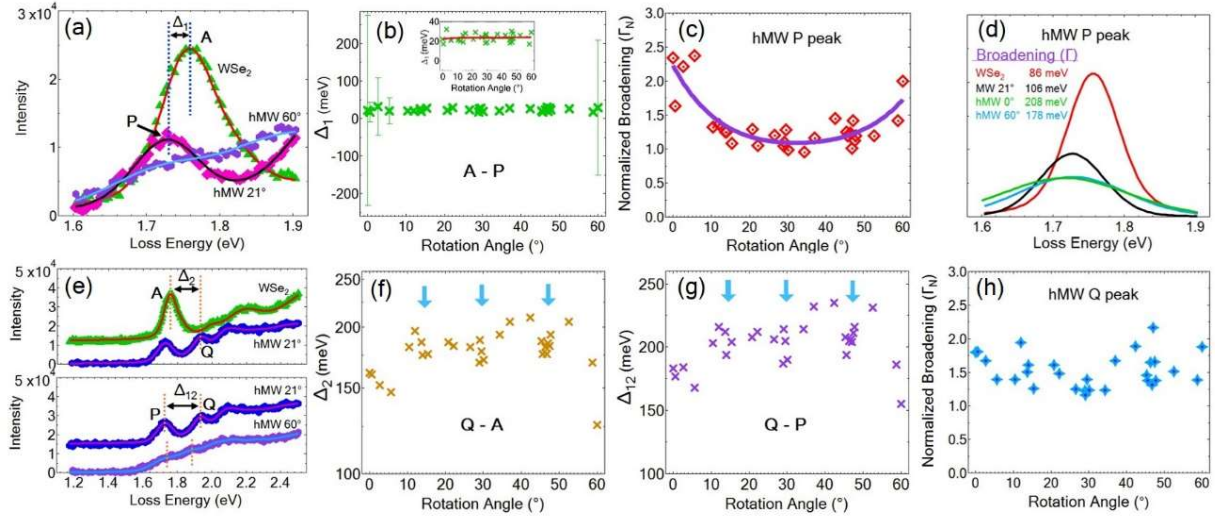

**Figure 3.** Peak-position and broadening of the P, Q peaks of hMW and their comparison to the A peak of monolayer $WSe_2$. (a) Comparison of the lineshape of the P peak of hMW with 21° and 60° with that of the A peak of monolayer $WSe_2$. (b) Plot of $\Delta_1$ as a function of rotation angle. (c) Plot of the normalized broadening ($\Gamma_N$) of the P peak with respect to the broadening of the A peak of monolayer $WSe_2$ as a function of rotation angle. (d) Plot of the lineshape of the Gaussian-Lorentzian oscillator corresponding to the A peak of $WSe_2$ and P peak of hMWs with 21°, 0°, and 60° showing their respective broadening ($\Gamma$). (e) Top panel shows $\Delta_2$, the difference of the peak position between the A peak of $WSe_2$ and the Q peak of hMW (with 21°). Bottom panel shows $\Delta_{12}$, the peak position difference between the P and Q peaks of hMW (with 21° and 60°). (f) and (g) are plots of $\Delta_2$ and $\Delta_{12}$, respectively, as functions of rotation angle. (h) Plot of the normalized broadening ($\Gamma_N$) of the Q peak with respect to the broadening of the A peak of monolayer $WSe_2$ as a function of rotation angle.

We also observe a distinct shift of the Q peak of hMW for different rotation angles (Figure 3(e)). To quantify this shift, we plot both $\Delta_2 = Q - A$, the energy difference of the Q peak of hMW with respect to the A peak of WSe$_2$, as well as $\Delta_{12} = Q - P$, the energy difference of the P and Q peaks of hMW, in Figure 3(f) and Figure 3(g), respectively. Intriguingly, both $\Delta_2$ and $\Delta_{12}$ show variations in the range of $50 \pm 10$ meV which are much larger than for $\Delta_1$. Also, we observe weak signatures of interesting dips in the plots of $\Delta_2$ and $\Delta_{12}$ near 15°, 30°, and 45° (indicated by ↓). We further note that $\Delta_{12}$ is the sum of $\Delta_1$ and $\Delta_2$. As $\Delta_1$ is more or less constant, $\Delta_{12}$ appears to be a trivial offset of $\Delta_2$. Hence, $\Delta_2$ depends almost linearly on the shift of the Q peak of hMW with respect to the A peak of MoS$_2$. As we interpret the Q peak of hMW as the shifted intralayer A peak of MoS$_2$ due to screening, it is inferred that the screening effect is larger on the excitonic absorption of MoS$_2$ in comparison to WSe$_2$. However, the broadening of the Q peak of hMW shows nontrivial rotation angle dependence with no distinct trend (unlike the P peak) as can be seen from Figure 3(h). We propose this to be a result of the different band arrangements and available diverse relaxation pathways in the valance bands of hMW.[7, 46, 47]

**Ultrafast charge transfer induced broadening of P peak**

To understand the important observation of the rotation-angle-dependent broadening of the P peak, we look deeper into the charge carrier dynamics process during the intra- and interlayer exciton formation due to excitation of quasiparticles. Due to the work function difference and the different positions of the valence and conduction bands, the hMW band structure is known to form a staggered type-II band alignment as shown in Figure 4(a) (see Supplementary Figure S10). Favourable energetics pushes the excited electron from the conduction band of $WSe_2$ to the conduction band of $MoS_2$, while the hole in the valence band of $MoS_2$ moves to the valence band of $WSe_2$. These electrons and holes in separate layers form interlayer excitons. The charge transfer occurs extremely fast within a timescale of 20 - 50 fs for the $MoS_2/WS_2$ heterostructure.[12, 30, 31] One distinct features of these heterostructures is the momentum space indirect character, which can be tuned by changing the rotation angle of the layers. This feature is shown schematically in Figure 4(b) for hMW, where the K (K′) valleys of the Brillouin zones of $MoS_2$ and $WSe_2$ in the conduction and valence bands, respectively, are separated by a finite momentum difference for misaligned cases. However, for the aligned and anti-aligned cases, these valleys become direct in nature, which dramatically enhances the charge transfer process leading to the efficient formation of the interlayer exciton.

As shown in Figure 4(c), the broadening of the A peak of $WSe_2$ has various contributions. $\Gamma_{exp}$ is the experimental broadening due to the energy spread and instability of the electron beam, spectrometer aberrations etc. (30 $\pm$ 5 meV), $\Gamma_{rad+phonon}$ is the radiative recombination lifetime and phonon assisted intra and intervalley scattering (in the same layer) induced broadening (~40 meV), and $\Gamma_{dop+inhomo}$ is the doping and inhomogeneity induced broadening (~10 meV).[12, 44]

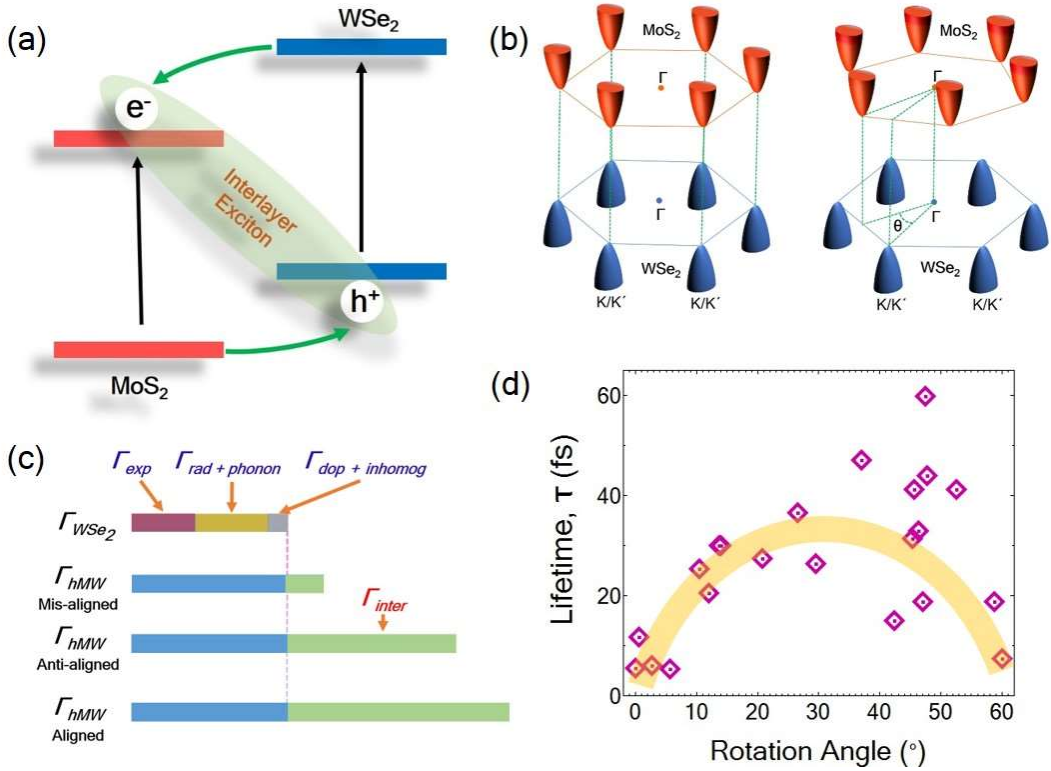

**Figure 4.** Schematic of the interlayer exciton formation process and the dependence of the lifetime of the excited carriers on the relative alignment of the Brillouin-zones of $MoS_2$ and $WSe_2$. (a) Interlayer exciton formation between the respective K-valleys of $MoS_2$ and $WSe_2$, after the transfer of the excited electrons in the conduction bands (from $WSe_2$ to $MoS_2$) and holes in the valence bands (from $MoS_2$ to $WSe_2$). (b) Schematic of the relative positions of the valence band (from $WSe_2$) K (K′) valleys and conduction band (from $MoS_2$) K (K′) valleys for aligned (or ant-aligned) and misaligned orientations. (c) Comparison of the broadening of the A peak of monolayer $WSe_2$ with that of the broadening of the P peak of hMW in the misaligned, anti-aligned, and aligned cases. (c) Plot of τ, the lifetime of the exciton corresponding to the P peak of hMW as a function of the rotation angle.

These contributions add up consistently to the observed broadening of $(80 - 89) \pm 5$ meV of the A peak of all the WSe$_2$ samples probed in this work. For the hMW, the P peak originates from the A peak of WSe$_2$, so inherently, it has the broadening contributions of the isolated monolayer WSe$_2$. However, the critical difference is the additional broadening channel introduced due to the interlayer charge transfer (or scattering) to the momentum mismatched (or matched) K (K′) valley in the conduction band of MoS$_2$. This additional broadening can dominate the aligned and anti-aligned cases (or cases very close to those) (Figure 4(c)). An estimate of the broadening contribution due to this ultrafast interlayer charge transfer process can be obtained to a good approximation by subtracting the A peak broadening of WSe$_2$ from the broadening of the P peak of hMW.[12] Using Heisenberg's uncertainty relation, the population lifetime, $\tau$ is calculated from this observed broadening, $\Gamma_{inter}$ due to the interlayer charge transfer process as $\tau = \hbar/\Gamma_{inter}$. Since, the experimental error bar in the estimation of $\Gamma_{inter}$ is 10 meV, which is equivalent to 60 fs, we plot the lifetime $\tau$ until 60 fs as a function of the rotation angle in Figure 4(d) (see Supplementary Figure S9). Remarkably, we observe the variation of $\tau$ in the range of 5 fs to 60 fs, where the smallest values of 5 fs are found for the aligned and anti-aligned cases or angles very close to these. This clearly demonstrates that momentum conservation critically enhances the charge transfer dynamics by about one order-of-magnitude, far beyond that reported previously for interlayer excitons in similar systems.[12, 30]

**Temperature-dependent EEL spectra**

To unveil the role of phonon scattering, lattice expansion, and the increase of interlayer separation on the excitonic features and their dynamics in hMW, we further perform temperature-dependent EEL measurements on a typical misaligned hMW from room temperature

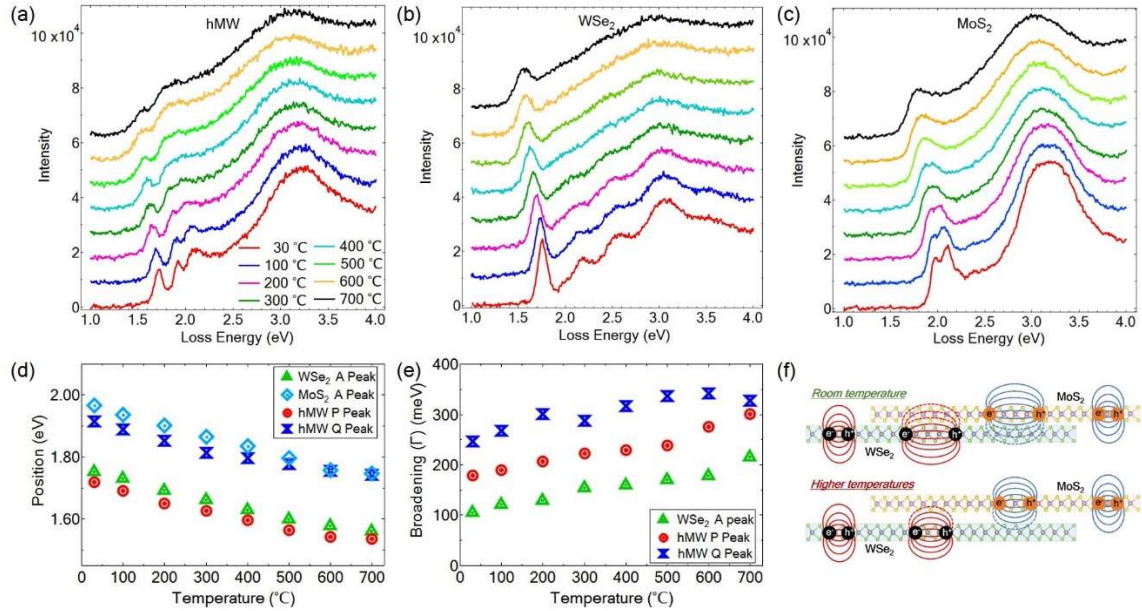

**Figure 5.** Temperature dependence of the EEL spectra. (a), (b), and (c) represent the temperature-dependent EEL spectra of a misaligned hMW, monolayer WSe$_2$, and monolayer MoS$_2$, respectively, from room temperature to 700 ˚C. (d) Temperature dependence of the peak positions of monolayer WSe$_2$, monolayer MoS$_2$, and hMW. (e) Temperature dependence of the broadening of the A peak of monolayer WSe$_2$ and P, Q peaks of hMW. For clarity, the P and Q peak broadenings are offset by 70 meV and 140 meV, respectively. (f) Schematic of the increase of the interlayer distance at higher temperatures, and its effect on the mutual interlayer screening effect of MoS$_2$ and WSe$_2$ layers in hMW. Increase of the interlayer distance at higher temperature reduces the screening.

(30 ºC) to 700 ºC, and compare the results with those obtained from monolayer WSe$_2$ and MoS$_2$. The EEL spectra for hMW, WSe$_2$, and MoS$_2$ are shown in Figure 5(a), (b), and (c), respectively. A general redshift is observed for all the peaks at higher temperatures, which can be understood as due to lattice renormalization.[48-50] Moreover, the features are broadened at higher temperatures due to enhanced phonon scattering. In Figure 5(d), the positions of the A peaks of $WSe_2$ and $MoS_2$, and the P, Q peaks of hMW are plotted as a function of temperature. With increasing temperature, the trend of shift of the P and Q peaks of hMW are similar to the A peaks of $WSe_2$ and $MoS_2$, respectively. This supports the hypothesis that the P, Q peaks are intrinsically due to intralayer absorption, but redshifted by enhanced screening from the adjacent layer. We note that even at the highest temperature of 700 ºC, the P peak shows a screening mediated redshift, which eliminates possibilities of this peak originating from other processes such as trion formation due to charge transfer. As schematically shown in Figure 5(f), higher temperature increases the interlayer distance by fractions of an angstrom, as observed in $MoSe_2$ for a similar range of temperature increase,[51] and hence the screening effect is observable to up to 700 ºC for $WSe_2$. However, we observe the Q peak position merging with the A peak position of $MoS_2$ from about 500 ºC, which is non-trivial to understand. This is potentially due to the different excitonic bonding strength and the dynamical nature of the constituent elements at different temperatures ($MoS_2$ atoms are lighter than $WSe_2$), but a precise understanding is beyond the scope of this work. Overall, the screening modulation of the Q peak with rotation angle of hMW is most likely due to steric effects, where the interlayer distance changes with rotation angle due to the finite and unequal sizes of the constituent atoms. Such steric effects are responsible for the rotation-dependent band-gap modulation in $MoS_2$ homobilayers, and for the local modulation of the quasiparticle gap in heterostructures as revealed by scanning tunneling spectroscopy.[10, 21, 22, 39] The broadening of the A peak of $WSe_2$, and P, Q peaks of hMW as a function of temperature are plotted in Figure 5(e), which show that the phonon scattering effects are similar for the three cases. The Q peak has slightly more variations, but otherwise these trends support the hypothesis that the P, Q peaks of hMW are intralayer absorption peaks. It is also noted that, for the aligned (or anti-aligned) case, the broadening of the P peak is affected by interlayer scattering to almost the same extent as for thermal phonon scattering at 700 ºC, where the broadening reaches ~200 meV.

We note that the broadening of the A peak of monolayer $WSe_2$ can be tuned using strain.[43, 44] This strain tuning are possible because the $WSe_2$ conduction band Q valley minimum is almost degenerate in energy with the K valley minimum. Application of positive or negative strain can change the position of the Q valley minimum, which can help or hinder phonon-assisted scattering of the excited electrons in the conduction band. Thus the A peak becomes broadened or sharpened with strain. Interestingly, various DFT based bandstructure calculations of vdW heterostructures also incorporate such strain effects to various degrees to obtain manageable supercell sizes for computational feasibility [7, 46, 52]. Such bandstructure calculations of hMW reveal that the valence band in the Γ valley is lifted considerably and the states in the conduction band Q valley are also affected.[7, 52] These changes are seen due to interlayer interaction (hybridization) and strain. However, there is no clear consensus on the amount of strain in a real sample and the effects seen in these theoretical results could be affected by the strain introduced artificially.[46] Moreover, Kunstmann *et al.* has shown that the conduction band Q valley rather shifts upward when the interlayer distance decreases.[7] These observations indicate that the potential alternative mechanism of the broadening of the P peak of hMW, as we observe experimentally, due to phonon-assisted intralayer and intervalley scattering of electrons is inadequate.

## Conclusion

In summary, we have demonstrated that rotational alignment of the constituent layers plays a dominant role in the interlayer charge transfer process, which is responsible for the formation of the interlayer exciton in vdW heterostructures. There exists an additional relaxation channel for the excited electrons and holes due to interlayer charge scattering, and this manifests as enhanced broadening in the hMW peaks. We estimate the fastest lifetime in the aligned (anti- aligned) case to be about 5 fs, which is one order-of-magnitude faster than the nominal lifetime of ~60 fs for the hMW with largest rotational mismatch (~30˚). Our results underline the role of the fundamental principle of momentum conservation in the interlayer exciton dynamics and provide a definitive guideline for the efficient design of functional devices harnessing interlayer excitons.

## Methods

**STEM-EEL Spectroscopy**

The electron microscope used for imaging and spectroscopy in this work is a JEOL STEM system (3C2), which is operated at the accelerating voltage of 60 kV. This system is equipped with Schottky field emission gun, a double Wein filter monochromator, and delta correctors. EEL spectra, acquired in the STEM mode, are collected by setting the energy resolution to 30 meV at FWHM of the zero-loss peak. The dispersion used is 5 meV/channel. The convergence semiangle $\alpha$ is 40 mrad and the collection semiangle $\beta$ is 125 mrad. The probe current is 10 pA for the EEL spectroscopy. EEL spectra are acquired in the dual EELS mode to eliminate any systematic error due to zero-loss shift.

For imaging, HAADF detector is used in the same STEM system for each sample location (with different relative orientation of the $MoS_2$ and $WSe_2$ layers, as well as for monolayer $MoS_2$ and $WSe_2$ regions), immediately after or before the EEL spectra is acquired. The convergence and collection semiangles are 40 mrad and 65 mrad, respectively.

**Chemical Vapour Deposition of $MoS_2$**

The $MoS_2$ monolayer, with triangular shaped single crystals (~1-3 micron sides) with high coverage, is deposited on commercially bought c-plane (0001) sapphire ($Al_2O_3$) substrate (Namiki Inc) in a quartz tube furnace (planarTECH LLC) with two separate upstream and downstream heating zones. In the downstream region, 4.7 mg of $MoO_3$ (99.98%, Sigma Aldrich) is placed in a single open-end crucible with a piece of nickel foam (size 3.5 cm x 3 cm, 1 mm thickness with 400 μm average pore size) placed directly above the $MoO_3$ powder. The sapphire is placed above the nickel foam, supported by two ceramic pieces. The arrangement of the $MoO_3$ crucible, nickel foam and sapphire is similar to the work of Lim *et al.*[53] 960 mg of sulfur (99.998%, Sigma Aldrich) is placed in the upstream zone, 40 cm away from the $MoO_3$ precursor. The detailed process flow is as such: the downstream region is first heated at 250 $^0$C for 10 minutes for degassing purposes with the upstream zone kept at room temperature while an Ar flow of 200 sccm is maintained. The temperature is then ramped to 750 $^0$C over 12 minutes and is maintained for an additional 10 minutes with an Ar flow of 50 sccm. Following which, the tube is allowed to naturally cool down to 650 $^0$C prior to opening of the furnace heaters for rapid cooling through ambient exposure while the Ar flow is increased to 200 sccm. Pressure is maintained at 6 Torr during the entire process.

**Chemical Vapour Deposition of WSe$_2$**

The WSe$_2$ monolayer, with larger triangular single crystals (~10 micron sides) and high coverage, is deposited on sapphire substrate using CVD as reported in the work of Huang *et al*.[52] High purity Se and WO$_3$ powders are used as the precursors, where the selenization of WO$_3$ is activated by the introduction of hydrogen in the reaction chamber. Ar is used as the carrier gas, where the WO$_3$ is placed in the center of the furnace on a ceramic boat. The sapphire target substrate is positioned in the downstream side, while the ceramic boat containing Se is placed in the upstream side. For the growth of larger isolated single layers, temperature is maintained at 850 $^0$C.

**Sample transfer to TEM grid**

The hMW samples on quantifoil grids are prepared by a double transfer method. First, the MoS$_2$ sample is spincoated with polycarbonate and then, with the help of this polycarbonate support film, the sapphire substrate is detached from the MoS$_2$ film in a dilute HF solution. This film is scooped using the sapphire substrate with the WSe$_2$ monolayer on top. After spincoating polycarbonate for the second time, the sapphire substrate is detached in dilute HF solution like before. The floating heterostructure sample with polycarbonate support film is transferred to beakers with ultrapure water consecutively to get rid of HF residue. Finally, the sample is fetched using the quantifoil grid, and the polycarbonate support film is dissolved in chloroform. After dipping in acetone and isopropyl alcohol, the sample is dried in vacuum.

ASSOCIATED CONTENT

The following Supplementary Information file is available free of charge.

File name: *Gogoi_hMW_ACSNANO_SI*

**Contents**

I. Raw EEL spectra of hMW, WSe$_2$ and MoS$_2$

II. Zero- loss subtraction

III. Normalization procedure, and normalized spectra for all 29 hMW

IV. Representative locations (holes) and HAADF images

V. Variations of the EEL spectrum of single layer WSe2 in different holes and near cuts in the same hole

VI. Moiré pattern difference between hMW 0° and hMW 60°

VII. Composite weighted average EEL spectra

VIII. Comparison with loss-function and $\epsilon_2$ of MoS$_2$ and WSe$_2$ from optical measurements (ellipsometry and reflectivity)

IX. Lifetime for all samples

X. Bandstructure of hMW

AUTHOR CONTRIBUTIONS

K.S. conceived the project. L.Y.C. and P.K.G prepared the TEM grids. P.K.G., R.S., and L.Y.C. performed the STEM imaging and EEL spectroscopy. P.K.G. analyzed the data. S.L.W. performed CVD growth of $MoS_2$ samples. D.C. supervised $MoS_2$ growth. L.J.L. supervised $WSe_2$ CVD growth. H.P.K. and A.V.K. provided theoretical inputs and computational results. All authors discussed the results and related interpretation. P.K.G. wrote the manuscript with contributions from K.S., L.Y.C, and R.S.

ACKNOWLEDGEMENTS

This work is (partially) supported by JSPS KAKENHI (16H06333, 17H04797, 18K14119, and P18350). P.K.G., M.B.H.B., and S.J.P. acknowledge MOE, Singapore grant number R-144-000-389-114. S.L.W. and D.C. acknowledge support from the Institute of Materials Research and Engineering (IMRE) under the Agency for Science, Technology, and Research (A*STAR) under A*STAR Science and Engineering Research Council Pharos 2D Program (SERC Grant No 152-70-00012). H.P.K. and A.V.K. thank CSC–IT Center for Science Ltd. for the computing resources.